\documentclass[12pt]{iopart}
\usepackage{epsfig,float,iopams,setstack}

\DeclareMathAlphabet{\mathsfsl}{OT1}{cmss}{m}{sl}
\begin{document}
\title[]{\bf \Large Dirac Analysis and Integrability of Geodesic Equations
for Cylindrically Symmetric Spacetimes}
\author{Ugur Camci}
\address{Department of Physics, Art and Science Faculty,
\d{C}anakkale Onsekiz Mart University, 17100 \d{C}anakkale,
Turkey}

\ead{ucamci@comu.edu.tr}

\begin{abstract}
Dirac's constraint analysis and the symplectic structure of
geodesic equations are obtained for the general cylindrically
symmetric stationary spacetime. For this metric, using the
obtained first order Lagrangian, the geodesic equations of motion
are integrated, and found some solutions for Lewis, Levi-Civita,
and Van Stockum spacetimes.
\end{abstract}


\bigskip

\qquad \qquad \today

\section{Introduction}    

Dirac first worked out the theory of quantizing constrained
systems in general\cite{dirac1}, and general relativity in
particular\cite{dirac2}, and his pioneering work continues to
serve as the foundation of current efforts to canonically quantize
gravity. Besides the hamiltonian formalism utilizing the
Schr\"odinger representation, there is an alternative Hamiltonian
approach which is based on Dirac's analysis\cite{dirac3} of
constrained systems. Many more extensive studies of the subject
can be found in the literature; see, for example, Refs. 4 and 5.
In this context, we discuss Dirac analysis of geodesic equations
for the general cylindrically symmetric stationary spacetimes, and
later integrate the obtained first order equations of motion for
these spacetimes.

Because of both the mathematical simplicity and the physical
relevance to our realistic world\cite{bonnor1,islam}, space-times
with cylindrical symmetry have been extensively studied, and their
relativistic applications have been further discussed recently
\cite{bonnor2}$^{-}$\cite{qadir}. The general form of this metric
in vacuum case was given by Lewis\cite{lewis}. Lewis stationary
vacuum metric is usually presented with four
parameters\cite{kramer} which admits a specific physical
interpretation when matched to a particular source. These four
parameters which are related to topological
defects\cite{silva1,herrera} not entering into the expression of
the physical components of curvature tensor may be real (Weyl
class) or complex (Lewis class). In recent years, the physical
meaning of these parameters have been discussed for both
classes\cite{silva1,silva2}. The corresponding static limit of the
the Lewis class was obtained by Levi-Civita (LC)\cite{lc}. Even in
the simplest case of the LC solution, its physical interpretation
is not completely understood, yet. In general, it contains two
independent parameters, in which there is only one mass parameter.
One of them is associated with the topological defects while the
second parameter is connected with the mass per unit length.
Another special case of Lewis metric is the van Stockum
solution\cite{vanst} which represents the gravitational field
produced by a rigidly rotating dust cylinder with a finite
thickness. The matching of this space-time to the vacuum Lewis
space-time  was also completed in Ref.$12$, and studied in detail
by Bonnor\cite{bonnor3}.

    The paper is organized as follows. In the next section we
present  the stationary cylindrical spacetime in general and give
the familiar spacetimes at the exterior of the boundary of the
source. In Sec. \ref{DA}, we shall present the symplectic
structure of geodesic equations for the general cylindrically
symmetric stationary metric. For the first order form of the
Lagrangian that yields the geodesic equations for this metric, we
shall apply Dirac's theory of constraints\cite{dirac3} to the
degenerate the Lagrangian. We find that the constraints are second
class as in the case of all integrable systems. The constraint
analysis yields the Dirac brackets, or the Hamiltonian operators
in the language of integrable systems. The symplectic $2$-form is
obtained by the Poison bracket of Dirac's constraints which is
also the inverse of the Hamiltonian operator. In Sec. \ref{int},
using the first order Lagrangian for the general cylindrically
symmetric metric, the geodesic equations are integrated, and the
obtained results are worked in cases of the Lewis spacetime for
the Weyl class, the exterior van Stockum and LC spacetimes.

\section{Spacetime}
\label{ST}

  The general line element for a cylindrically symmetric stationary
spacetime is given by
\begin{equation}
ds^2 = -fdt^2 + 2k dt d\phi + e^{\mu}(dr^2 + dz^2) + \ell d\phi^2,
\label{met}
\end{equation}
where $f, k, \mu$ and $\ell$ are functions only of $r$, and
$x^i=(t,r,z,\phi,t), i = 0,1,2,3$ are the usual cylindrical
coordinates with
\begin{equation}
-\infty \le t,z \le \infty,\,\,\,\, r \geq 0,\,\,\,\, 0 \leq \phi
\leq 2\pi
\end{equation}
and the hypersurfaces $\phi = 0, 2\pi$ being identified.
Einstein's field equations for vacuum are
\begin{equation}
R_{i\,j} = 0. \label{vac}
\end{equation}
The general solution of (\ref{vac}) for (\ref{met}) is the
stationary Lewis metric\cite{lewis}, which can be written as
\cite{kramer}
\begin{eqnarray}
f &=& ar^{-n+1} -\frac{c^2}{n^2 a} r^{n+1}, \label{lewis1} \\ k
&=& -A\,f, \label{lewis2}
\\ \ell &=& \frac{r^2}{f} - A^2 f, \label{lewis3} \\ e^{\mu} &=&
r^{\frac{1}{2}(n^2-1)} \label{lewis4}
\end{eqnarray}
with
\begin{equation}
A = \frac{c\,r^{n+1}}{n a f} + b.
\end{equation}
The constants $n, a, b$ and $c$ can be either real or complex, the
corresponding solutions belong to the Weyl or Lewis classes,
respectively. For the Weyl class, the above parameters have the
following physical interpretations. The parameter $n$ is
associated  with the Newtonian mass per unit length of an uniform
line mass $\sigma$ when it produces the low density regime. The
parameter $a$ is related to the constant arbitrary potential that
exist in the corresponding Newtonian solution, while the
parameters $b$ and $c$ are responsible for the non-staticity of
the spacetime, since when we take $b = 0$ and $c = 0$ the Weyl
class reduces to the static LC metric. The parameter $b$ is
related, in locally flat limit, with the angular momentum of a
spinning string. The parameter $c$ measures the vorticity of the
source when it represented by a stationary completely anisotropic
fluid. For further details see Ref. $9$. For the Lewis class, the
physical and geometrical meaning of the four parameters of the
Lewis metric was given in Ref. $10$.

    For the line element of the LC static vacuum spacetime in the Weyl form,
the functions $f, k, \mu$, and $\ell$ have the following
expressions\cite{bonnor1,silva1,wang}
\begin{equation}
f = r^{4 \sigma} \quad k = 0, \quad \ell = C^{-2} r^{2-4\sigma},
\quad e^{\mu} =  r^{4\sigma (2\sigma -1)} \label{lc}
\end{equation}
where $\sigma$ and $C$ are two arbitrary constants and both of
them are fixed by  the internal composition of the physical
source. The constant $C$ refers to the angular
defect\cite{silva1}, and cannot be removed by scale
transformation. This constant is related, in the locally flat
limit, with the parameter $a$ in the Lewis solution for the Weyl
class, given by
\begin{equation}
a = C^2. \label{a-C}
\end{equation}
The physical importance of the other parameter $\sigma$ is mostly
understood in accordance with the Newtonian analogy of the LC
solution, i.e. the parameter $\sigma$ represents the mass per unit
length\cite{silva1,wang}. The parameter $\sigma$ is connected to
the parameter $n$ in the Lewis spacetime for the Weyl class as
\begin{equation}
n = 1 - 4\sigma.
\end{equation}

    The functions $f, k, \mu$, and $\ell$ for the van Stockum
solution\cite{vanst} are given by
\begin{equation}
f = 1,\quad  k= \alpha r^2, \quad \ell = r^2(1-\alpha^2 r^2),
\quad \mu = -\alpha^2 r^2 \label{vanStockum}
\end{equation}
with $\alpha$ being an arbitrary positive constant. The energy
density and the four velocity of the dust are $$ \rho =
\frac{\alpha^2}{2\pi G} e^{\alpha^2 r^2}, \quad u^{\mu} =
\delta_4^{\mu}$$ where $G$ is the gravitational constant. The
angular velocity to the fluid with respect to a locally
nonrotating frame  is $ \omega = \alpha (1-\alpha^2 r^2 )^{-1}$.
Since near the axis, $\omega \rightarrow \alpha$, it can be
interpreted that $\alpha$ is the angular velocity of the fluid on
the axis\cite{bonnor3}.

    The van Stockum exterior solution (\ref{vanStockum}), which is
a particular case of the Lewis metric, contains the globally
Minkowski spacetime as a special case. Therefore the van Stockum
solution (\ref{vanStockum}) must be a particular case of the Weyl
class\cite{silva1}. Since the van Stockum spacetime cannot be
reduced to the globally static LC metric (\ref{lc}), it is a
particular case of the Weyl class with $b \neq 0$ and $c \neq 0$,
since for $b = 0$ and $c = 0$ the Weyl class can be globally
reduced to the static LC metric (\ref{lc}).

\section{Dirac Analysis for Geodesic Equations}
\label{DA}

        The equations governing the geodesics can be derived from the
lagrangian
\begin{equation}
2{\cal L} = g_{i\, j} \frac{dx^i}{d\tau} \frac{dx^j}{d\tau}
\label{geolag}
\end{equation}
where $\tau$ is an affine parameter along the geodesics. From the
external problem it emerges the Euler-Lagrange equations
\begin{equation}
\frac{d}{d\tau} \left(\frac{\partial {\cal L}}{\partial \dot{x}^i}
\right) - \frac{\partial {\cal L}}{\partial x^i} = 0
\end{equation}
and from them follow the geodesics given by
\begin{equation}
\ddot{x}^i + \Gamma^i_{j\,k} \dot{x}^j \dot{x}^k = 0 \label{geoeq}
\end{equation}
where the overdot denotes differentiation with respect to $\tau$.
For spacetime (\ref{met}) the Lagrangian (\ref{geolag}) is
\begin{equation}
{\cal L}_L = \frac{1}{2} f \, \dot{t}^2 -k \dot{t}\dot{\phi}
-\frac{1}{2}e^{\mu} (\dot{r}^2 + \dot{z}^2) - \frac{1}{2} \ell
\dot{\phi}^2 . \label{lag1l}
\end{equation}
This Lagrangian is second order and therefore not suitable to a
discussion of symplectic structure. For purposes of Hamiltonian
analysis we need to start with first order Lagrangian and it can
be verified that
\begin{eqnarray}
{\cal L} &=&-\frac{1}{2\ell} P \, Q + \frac{1}{2\ell} \Big{[}
k(P+Q)-D(P-Q)
\Big{]} \dot{t} + \frac{1}{2} (P+Q) \dot{\phi} \nonumber \\
& &-\frac{1}{2}e^{-\mu}(M^2+N^2) + M \dot{r} + N \dot{z}
\label{lag2}
\end{eqnarray}
gives rise to the equations of motion
\begin{eqnarray}
& & P = \ell \dot{\phi} + (k+D) \dot{t},\,\,\, Q = \ell \dot{\phi}
+ (k-D) \dot{t}, \label{eqm1} \\& & M = e^{\mu} \dot{r},\qquad  N
= e^{\mu} \dot{z}, \label{eqm2} \\& & \dot{M} -
\frac{1}{2}e^{-\mu} \mu' (M^2+N^2) - \frac{\ell'}{2\ell^2} P\,Q -
\frac{(P - Q)} {4D} F = 0, \label{eqm3}
\end{eqnarray}
\begin{eqnarray}
& & \left[ \frac{k(P+Q)-D(P-Q)}{\ell} \right]^{\bf .} = 0,
\label{eqm4_1} \\& &  \dot{P} + \dot{Q} = 0, \label{eqm4_2} \\& &
\dot{N} = 0, \label{eqm4_3}
\end{eqnarray}
which together result equations (\ref{scndord1}-\ref{scndord3}) in
\ref{apa}, where we have defined $$D^2 \equiv k^2 + \ell f$$ and
$$F \equiv(P+Q)\left(\frac{k}{\ell}\right)^{\prime}-(P-Q)\left(\frac{D}{\ell}
\right)^{\prime},$$ and the prime represents derivative with
respect to $r$. Dirac quantization is canonically quantize the
original phase space which is usually even dimensional symplectic
manifold, and then imposed the gauge constraints as operator
conditions on the physical quantum states. In this first order
formulation we have introduced $x^a \equiv (P,Q,M,N), \, a=
4,5,6,7,$ as new variables which is double the number required. So
we consider a symplectic manifold spanned by variables $X^A=
(x^i,x^a), A = 0,...,7,$ where $x^i$'s and $x^a$'s are,
respectively, spacetime and configuration space variables. Then,
first order field equations become
\begin{equation}
\dot{X}^A = {\bf X}\,(X^A),
\end{equation}
with the vector field defining the flow
\begin{eqnarray}
{\bf X}&=& \frac{1}{2 D} (P - Q) \frac{\delta}{\delta{t}} +
e^{-\mu} \left( M \frac{\delta}{\delta{r}} + N
\frac{\delta}{\delta{z}} \right) \nonumber \\& &  + \frac{1}{2
\ell} \left[ P + Q - \frac{k}{D} (P - Q) \right]
\frac{\delta}{\delta{\phi}} + \frac{M \ell e^{-\mu}}{2D} F \left(
\frac{\delta}{\delta{P}} - \frac{\delta}{\delta{Q}} \right)
\nonumber \\& & + \Bigg{[} \frac{1}{2}e^{-\mu} \mu' (M^2 + N^2) +
\frac{\ell'}{2\ell^2} P\,Q + \frac{(P - Q)}{4D} F \Bigg{]}
\frac{\delta}{\delta{M}} \label{vectorf}
\end{eqnarray}
for the geodesic equation (\ref{geoeq}).

The Lagrangian (\ref{lag2}) is degenerate because its Hessian
\begin{equation}
\det \left| \frac{\partial^2 {\cal L}}{\partial \dot{X}^A \,
\partial \dot{X}^B} \right| = 0 \label{hessian}
\end{equation}
vanishes identically. Hence it is a system subject to constraints
and the passage to its Hamiltonian structure requires the use of
Dirac's theory of constraints\cite{dirac3}. We introduce the
canonical momenta of the test particle defined by
\begin{equation}
\Pi_A\equiv\frac{\partial{\cal L}}{\partial{\dot{X}^A}}
\label{momenta}
\end{equation}
which cannot be inverted due to equation (\ref{hessian}). The
definition of the momenta therefore gives rise to the constraints
\begin{eqnarray}
\Phi_0 &=& \Pi_{t}-\left[
\frac{k}{2 \ell}(P + Q)-\frac{D}{2\ell}(P - Q) \right], \nonumber \\ \Phi_1 &=& \Pi_{r} - M, \nonumber \\
\Phi_2 &=& \Pi_{z} - N, \nonumber\\ \Phi_3 &=& \Pi_{\phi}-\frac{1}{2}(P + Q), \label{constraints}\\
\Phi_4 &=& \Pi_{P}, \nonumber \\ \Phi_5 &=& \Pi_{Q}, \nonumber\\
\Phi_6 &=& \Pi_{M}, \nonumber
\\ \Phi_7 &=&  \Pi_{N}, \nonumber
\end{eqnarray}
which must vanish weakly, i.e on shell. In order to determine the
class of these constraints we need to obtain the Poisson bracket
of the constraints
\begin{equation}
C_{AB}(\tau,\tilde{\tau})=\Big{\{}\Phi_A(\tau),\Phi_B(\tilde{\tau})\Big{\}}
\\
\end{equation}
using the canonical Poisson brackets
\begin{equation}
\Big{\{}X^A(\tau),\Pi_B(\tilde{\tau})\Big{\}}=\delta_B^A
\delta(\tau-\tilde{\tau})
\end{equation}
between the dynamical variables and their conjugate momenta. The
result
\begin{equation}
C_{AB}(\tau,\tilde{\tau})= \frac{1}{2} \left(
\begin{array}{cccccccc}
\, 0 &\, -F &\,\, 0 &\,\, 0 & \,\,\frac{-k + D}{\ell} &
\,\,-\frac{(k + D)}{\ell}& \,\,0 & \,\,0
\\ \, F &\,\,\, 0 &\,\, 0 &\,\, 0 & \,\,\,\,\,0 & \,\,\,\,\,0 &
-2 & \,\, 0 \\ \, 0 &\,\,\, 0 &\,\, 0 &\,\, 0 & \,\,\,\,\,0 &
\,\,\,\,\,0 & -2 & \,\, 0 \\ \, 0 &\,\,\, 0 &\,\, 0 &\,\, 0 &
\,\,\,-1 & \,\,\,-1 & \,\,\,0 & \,\, 0 \\ \, \frac{k - D}{\ell}
&\,\,\, 0 &\,\,0 &\,\, 1 & \,\,\,\,\,0 & \,\,\,\,\,0 & \,\,\,0 &
\,\,0 \\ \, \frac{k + D}{\ell} &\,\,\, 0 &\,\, 0 &\,\, 1 &
\,\,\,\,\,0 & \,\,\,\,\,0 & \,\,\,0 & \,\, 0 \\ \,\,0 &\,\,\,2
&\,\, 0 &\,\, 0 & \,\,\,\,\,0 & \,\,\,\,\,0 & \,\,\,0 & \,\, 0 \\
\, 0 &\,\,\, 0 &\,\, 2 &\,\, 0 & \,\,\,\,\, 0 & \,\,\,\,\, 0 &
\,\,\,0 & \,\,0
\end{array}
\right) \delta (\tau-\tilde{\tau}) \label{cmat}
\end{equation}
shows that the constraints (\ref{constraints}) are second class as
in the case of all integrable systems\cite{rma}.

   In order to obtain the Hamiltonian for the degenerate
Lagrangian (\ref{lag2}) we first construct the free Hamiltonian
obtained by Legendre transformation
\begin{eqnarray}
H_{0} & = & \Pi _{A}\dot{X}^{A}-{\cal L} \nonumber \\ &=&
\frac{1}{2\ell} P\,Q + \frac{1}{2} e^{-\mu} (M^2 + N^2) \label{h0}
\end{eqnarray}
and the total Hamiltonian density of Dirac is given by
\begin{equation}
 H_{T} = H_{0}+\lambda^{A}\Phi _{A}
\end{equation}
where $\lambda^A$ are Lagrange multipliers. Since we have second
class constraints the Lagrange multipliers will be determined from
the solution of
\begin{equation}
\Big{\{} H_T,\Phi_{A}\Big{\}}=0 \label{hold}
\end{equation}
which ensure that the constraints hold for all values of $\tau$.
Since the constraints are linear in the momenta the Lagrange
multipliers are given by
\begin{eqnarray}
\lambda^0 & = & \frac{1}{2D} (P - Q), \nonumber \\
\lambda^1 & = & M e^{-\mu},  \nonumber \\ \lambda^2 & = & N
e^{-\mu}, \nonumber \\ \lambda^3 & = &  \frac{1}{2\ell} (P + Q)
-\frac{k}{2\ell D} (P - Q), \\ \lambda^4 & = & \frac{M \, \ell e^{-\mu}}{2D} F,  \nonumber \\
\lambda^5 & = &  -\lambda^4 \nonumber
\\ \lambda^6 & = & \frac{\ell'}{2\ell^2} P\,Q + \frac{1}{2} e^{-\mu} \mu' (M^2 +
N^2) + \frac{F}{4D}(P-Q), \nonumber \\ \lambda^7 & = & 0 \nonumber
\end{eqnarray}
which follows directly from the flow (\ref{vectorf}).

   The Dirac bracket is a modification of the Poisson bracket designed
to vanish on the surface defined by the constraints. For two
smooth functionals ${\cal A},{\cal B}$ of the canonical variables
we have
\begin{equation}
\{ {\cal A} ,{\cal B} \}_D = \{ {\cal A},{\cal B} \} - \{ {\cal A}
, \Phi_A \} J^{AB} \{ \Phi_B, {\cal B} \} \label{diracbracket}
\end{equation}
where $J$ is obtained by inverting the matrix of the Poisson
bracket of the constraints $C$
\begin{equation}
\int C_{AB}(\tau, \tilde{\tilde{\tau}})
J^{BC}(\tilde{\tilde{\tau}}, \tilde{\tau}) \, d
\tilde{\tilde{\tau}} = \delta^C_A \, \delta(\tau-\tilde{\tau})
\label{invert}
\end{equation}
The inverse of the Poisson bracket of the constraints is known as
the \emph{Hamiltonian operator} in the literature of integrable
systems \cite{rma}. From Eq. (\ref{invert}) we obtain
\begin{equation}
J^{A\,B}(\tau,\tilde{\tau})=\frac{1}{2} \left(
\begin{array}{cccccccc}
0 &\,\,\, 0 &\,\, 0 &\, 0 & \,\, -\frac{\ell}{D}
&\,\,\,\frac{\ell}{D}  & \,\,\,0 & \,\,0 \\  0 &\,\,\, 0 & \,\,0
&\, 0 & \,\,\,\,0 & \,\,\,\,0 & \,\, 1& \,\,0
\\ 0 &\,\,\, 0 &\,\, 0 &\, 0
& \,\,\,\,0 & \,\,\,\,\, 0 & \,0 & \,\, 1 \\ 0 &\,\,\, 0
&\,\, 0 &\, 0 & \,\,\,\, \frac{k + D}{D} & \,\,\,\,\,\frac{-k +D}{D} & \,0 & \,\, 0
\\ 0 &\,\,\, 0 &\,\,0 &\, -\frac{(k+D)}{D} & \,\,\,\,0 &
\,\,\,\,\,0 & \,\,\,\frac{\ell F}{D} & \,0
\\ -\frac{\ell}{D} &\,\,\, 0 &\,\, 0 &\, \frac{k - D}{D} & \,\,\,\,0 & \,\,\,\,\,0 &
-\frac{\ell F}{D} & \,\, 0 \\  0 &\,\,\, -1 &\,\, 0 &\, 0 &
\,\, -\frac{\ell F}{D}& \,\,\, \frac{\ell F}{D} & \,0 & \,\, 0 \\
0 &\,\,\, 0 &\,\, -1 &\, 0 & \,\,\,0 & \,\,\,\,\, 0 & \,0 &
\,\,\,0
\end{array}
\right) \delta (\tau-\tilde{\tau}) \label{jmat}
\end{equation}
and Eqs.(\ref{eqm1})-(\ref{eqm4_3}) can be written in Hamiltonian
form
\begin{equation}
\dot{X}^A=J^{A\,B}\frac{\delta H_0}{\delta X^B}
\end{equation}
where integration over dotted variables is implied.

  The symplectic 2-form is given by \cite{rma}
\begin{equation}
\omega_D = \frac{1}{2} \delta{X^A} \, \wedge \,C_{A\,B}\,
\delta{X^B} \label{symp2def}
\end{equation}
and using Eq.(\ref{cmat}) we find that
\begin{eqnarray}
\omega_D &=&  \frac{1}{2} \delta{P} \wedge \delta{\phi} +
 \frac{1}{2} \delta{Q} \wedge \delta{\phi} + \delta M \wedge \delta r
 + \delta N \wedge \delta z + \frac{(k-D)}{2\ell} \delta{P} \wedge
 \delta{\phi} \nonumber  \\
& & +  \frac{(k+D)}{2\ell} \delta Q \wedge \delta t + \frac{1}{2}
\Big{[} (P+Q)\Big{(}\frac{k}{\ell}\Big{)}' -
(P-Q)\Big{(}\frac{D}{\ell} \Big{)}' \Big{]} \delta r \wedge \delta
t \label{symp2kD}
\end{eqnarray}

    Finally, Hamilton's equations can be written in the form
\begin{equation}
i_{\bf X}\omega_D=-\delta H_0
\end{equation}
where $i_{\bf X}$ denotes contraction with respect to the vector
field (\ref{vectorf}) of the symplectic $2$-form (\ref{symp2kD}).

  We shall now turn to the Witten-Zuckerman\cite{witten,zuck} formulation of
  symplectic $2$-form vector density $\omega$ which is closed
\begin{equation}
\delta \omega =0 \label{closed}
\end{equation}
and conserved
\begin{equation}
\dot{\omega}= 0. \label{conserved}
\end{equation}
Starting with the Lagrangian (\ref{lag2}) we find that the
Witten-Zuckerman symplectic $2$-form is the same expression for
the symplectic $2$-form (\ref{symp2kD}) obtained from Dirac's
theory of constraints, i.e. $$ \omega = \omega_D.$$

\section{Integration}
\label{int}

   The geodesic equations (\ref{geoeq}) are four equations for the
four unknowns $\dot{t}, \dot{r}, \dot{z}$ and $\dot{\phi}$ (see
\ref{apa}). Firstly, from the Eqs. (\ref{eqm1}) and (\ref{eqm2}),
we obtain
\begin{eqnarray}
\dot{t}&=& \frac{1}{2D}(P-Q), \label{soln1}
\\ \dot{r} &=& M e^{-\mu},\label{soln2}
\\ \dot{z} &=& N e^{-\mu}, \label{soln3} \\ \dot{\phi} &=&
\frac{1}{2\ell} \Big{[}(P+Q) -\frac{k}{D}(P-Q)
\Big{]}.\label{soln4}
\end{eqnarray}
Later, integrating Eqs.  (\ref{eqm4_1})-(\ref{eqm4_3}) we get
\begin{eqnarray}
N &=& K, \\ P + Q &=& 2 L, \\
\frac{D}{\ell}(P-Q)-\frac{k}{\ell}(P+Q) &=& 2 E,
\end{eqnarray}
where $K, L$, and $E$ are integration constants. Using these
results in Eq. (\ref{eqm3}) and rearranging, yields the following
Riccati type differential equation
\begin{equation}
\dot{M} = p (\tau) M^2 + q(\tau) \label{eqM}
\end{equation}
where $p$ and  $q$ are given by
\begin{eqnarray}
p &=& \frac{\mu'}{2} e^{-\mu}, \label{p} \\ q &=& \frac{1}{2} K^2
p + \frac{\ell'}{2\ell f} \left[ \Big{(} L f - E k \Big{)}^2 -E^2
D^2 \right] + \frac{(L k + E \ell )}{2 D^2} F, \label{q}
\end{eqnarray}
where $F \equiv 2 \left[ L \left( \frac{k}{\ell} \right)^{\prime}
- \frac{(L k + E \ell)}{D} \left(\frac{D}{\ell} \right)^{\prime}
\right]$. Then, the solution of Eq. (\ref{eqM}) is obtained as
\begin{equation}
M = \left(|\frac{q}{p}|\right)^{1/2}U \label{M}
\end{equation}
where the following integral equation must still be satisfied for
$U$,
\begin{equation}
\int{\frac{dU}{U^2 + \beta U +1}} + c_1 = \int{q
\left(|\frac{p}{q}|\right)^{1/2}} d\tau
\end{equation}
where $c_1$ is an integration constant, and $\beta$ is given by
\begin{equation}
\beta = \frac{1}{2p} \left(|\frac{q}{p}|\right)^{1/2} \left(
|\frac{p}{q}| \right)^{\bf .}
\end{equation}
Using (\ref{soln2}) in (\ref{M}), we find
\begin{equation}
\dot{r} = e^{-\mu} M. \label{rdot_son}
\end{equation}
Thus, we have obtained the generic expression for the radial speed
of the test particle. A detailed examination of the solutions of
the Eq. (\ref{eqM}) is beyond the scope of this paper;therefore,
we shall only consider some special orbits here. Let us first
assume that $p = -q = -1$. In this case, we find the following
solutions:
\begin{eqnarray}
& & e^{-\mu} = 2 \, r + m, \\& & r (\tau) = \left\{
\begin{array}{l}
\frac{1}{2} \left[ n^2 e^{\pm 2(\tau -\tau_0)} -m \right],\qquad
\quad \,\,\, for \, M = \pm1, \\
\frac{1}{2} \left[ n^2 \cosh^2(\tau- \tau_0) -m \right], \quad \,
for \,\, M= \tanh(\tau -\tau_0)  \\\frac{1}{2} \left[ n^2
\sinh^2(\tau- \tau_0) -m \right],  \quad  \, \, for \, M =
\coth(\tau - \tau_0)
\end{array}
\right.
\end{eqnarray}
where $m, n$, and $\tau_0$ are integration constants. Second, we
assume that $p = q = b \tau^n$, where $b$ and $n$ are nonzero
constants. In this case, it follows from Eqs. (\ref{eqM}) and
(\ref{p}) that for $n \neq -1$
\begin{eqnarray}
& & e^{-\mu} = a \cos^2 \left(\frac{b}{n+1} \tau^{n+1} - \tau_0
\right), \quad M = \tan\left( \frac{b}{n+1} \tau^{n+1} -\tau_0
\right), \\& & r (\tau) = \frac{a b \, sign(b)
\cos(2\tau_0)\tau^{2 + n}}{(1+n)(2+n) sign(1+n)} \nonumber \\& &
\qquad \quad \times\, _1 F_2 \left[ \{\frac{1}{2} +
\frac{1}{2(1+n)} \}, \{ \frac{3}{2}, \frac{3}{2} +
\frac{1}{2(1+n)} \}, - \frac{b^2\, \tau^{2(1+n)}}{(1+n)^2} \right]
\nonumber \\& & \qquad \quad - \frac{a \sin(2\tau_0) \tau}{2}  \,
_1 F_2 \left[ \{ \frac{1}{2(1+n)} \}, \{ \frac{1}{2}, 1 +
\frac{1}{2(1+n)} \}, - \frac{b^2\, \tau^{2(1+n)}}{(1+n)^2} \right]
\end{eqnarray}
and for $n = -1$
\begin{eqnarray}
& & e^{-\mu} = a \cos^2 \left[b \ln (\tau_0 \tau)\right], \quad M
= \tan [ b \ln(\tau_0 \tau) ],
\\& & r (\tau) = \frac{a\, \tau}{2(1+4b)^2} \cos[2 b \ln(\tau_0
\tau)]\left[ \tan[2 b \ln(\tau_0 \tau)] - 2 b \right],
\end{eqnarray} where $a$ and $\tau_0$ are constants of
integration, and $_1 F_2$ is the generalized hypergeometric
function.

    Considering the Eq. (\ref{momenta}), the momenta of the
test particle for metric (\ref{met}) using the Lagrangian
(\ref{lag2}) is given by
\begin{eqnarray}
\Pi_r &\equiv& M, \nonumber \\ \Pi_z &\equiv& K, \nonumber \\
\Pi_{\phi} &\equiv& \frac{1}{2}(P + Q) = L, \\ \Pi_t &\equiv&
\frac{1}{2\ell} \left[ k(P+Q) -D (P-Q) \right] = -E \nonumber
\end{eqnarray}
Hence $E$ can be interpreted as the total energy of the particle,
and will be always taken nonnegative. $K$ can be interpreted as
its momentum along $z$ and $L$ its angular momentum. In terms of
these conserved quantities, $\dot{t}, \dot{r}, \dot{z}$ and
$\dot{\phi}$ become
\begin{eqnarray}
\dot{t}&=& \frac{-(E \ell + L k)}{D^2} , \label{soln1_2}
\\ \dot{r} &=& \left(|\frac{q}{p}| \right)^{1/2} e^{-\mu} \, U,\label{soln2_2}
\\ \dot{z} &=& K e^{-\mu}, \label{soln3_2} \\ \dot{\phi} &=& \frac{E k -L
f}{D^2}. \label{soln4_2}
\end{eqnarray}

     As is well known, the Eq. (\ref{geoeq}) has a first
integral that is equivalent to $g_{\mu\nu}
\dot{x}^{\mu}\dot{x}^{\nu} = -\epsilon $, where $ \epsilon = 0,
1$, or $-1$ if the geodesics are respectively null, timelike or
spacelike. This implies that
\begin{equation}
-\epsilon = -f \dot{t}^2 + 2 k \dot{t}\dot{\phi} +
e^{\mu}(\dot{r}^2+ \dot{z}^2) + \ell \dot{\phi}^2
\label{firstint1}
\end{equation}
so that ${\cal L}_L$ in (\ref{lag1l}) is equal to
$\frac{\epsilon}{2}$ along the path. For the Lagrangian
(\ref{lag2}), this equation becomes
\begin{equation}
\frac{1}{\ell} P Q + e^{-\mu} (M^2 + N^2) = - \epsilon.
\label{firstint2}
\end{equation}
Now, substituting (\ref{soln1}), (\ref{soln3}) and (\ref{soln4})
into (\ref{firstint1}), we have an expression for the radial speed
$\dot{r}^2$ of the test particle :
\begin{equation}
\dot{r}^2 = e^{-\mu} \left[W_0 (r) -W(r) \right], \label{rdot1}
\end{equation}
where $W_0(r)$ and $W(r)$ are defined as
\begin{equation}
W_0 (r) = E^2 \frac{\ell}{D^2} + 2 E L \frac{k}{D^2} - \epsilon,
\qquad W (r) = L^2 \frac{f}{D^2} + K^2 e^{-\mu}.
\end{equation}
From  the Eqs. (\ref{soln2_2}) and (\ref{rdot1}), we find
\begin{equation}
U = \pm e^{\mu/2} \sqrt{\frac{p}{q}\left[ W_0 (r) -W (r)\right]}
\end{equation}
which enables us to relate Eqs. (\ref{soln2_2}) and (\ref{rdot1}).
On the other hand, setting $W_0 (r) = G(r)\left[V_0 -V_1
(r)\right]$ and $W = G(r) V_2 (r)$, then Eq.(\ref{rdot1}) becomes
\begin{equation}
\dot{r}^2 = e^{-\mu} G(r) \left[V_0 - V(r)\right], \label{rdot2}
\end{equation}
where $V = V_1 + V_2$. Then, one can obtain the functions $G(r)$
and $V(r)$, and the constant $V_0$ for spacetimes (\ref{met}) with
(\ref{lewis1})-(\ref{lewis4}), (\ref{lc}) and (\ref{vanStockum})
given in Sec. \ref{ST}.

    In the Lewis spacetime for the Weyl class,
which means that the parameters $n, a, b$ and $c$ appearing in
(\ref{lewis1})-(\ref{lewis4}) are real, we obtain $V_0, V(r)$ and
$G(r)$ as
\begin{eqnarray}
G(r) &=& r^{n-1}, \label{lw1}\\ V_0 &=& \frac{1}{a n^2} \left[(E b
+ L)c -E n \right]^2, \label{lw2}\\ V (r) &=& \epsilon r^{1-n} + a
(E b + L)^2 r^{-2n} + K^2 r^{(1-n)(3+n)/2}. \label{lw3}
\end{eqnarray}
In the general case, i.e. $b \neq 0$ and $c \neq o$, we see that
the parameter $c$ only effects $V_0$ by modifying the energy of
the test particle, and leaving otherwise the geodesics
indistinguishable from the static LC spacetime. In two different
subcases, when $b = 0, c \neq 0$ and $b \neq 0, c = 0$, the
physical interpretations are given by Herrera and Santos
\cite{herrera}. Furthermore, for the Lewis solution
(\ref{lewis1})-(\ref{lewis4}), the differential equation
(\ref{rdot2}) can be written in the integral form as
\begin{equation}
\int{\frac{r^{\frac{n-1}{\sqrt{2}}}
dr}{\sqrt{\frac{1}{a}\left[\Gamma \frac{c}{n} -E\right]^2 - a
\Gamma^2 r^{-2n} -\epsilon r^{1-n} - K^2 r^{(1-n)(3+n)}}}} = \pm
(\tau - \tau_0), \label{fint}
\end{equation}
where $\Gamma \equiv E b + L$, and $\tau_0$ is a constant of
integration. The $\pm$ signs in Eq. (\ref{fint}) correspond to
outgoing and ingoing geodesics, respectively.

    In the case of the LC solution (\ref{lc}), for $V_0, V(r)$ and
$G(r)$, we have
\begin{eqnarray}
G(r) &=& r^{-4\sigma}, \label{lc1} \\ V_0 &=& E^2, \label{lc2} \\
V (r) &=& \epsilon r^{4\sigma} + C^2 L^2 r^{2(4\sigma -1)} + K^2
r^{8 \sigma (1-\sigma)}. \label{lc3}
\end{eqnarray}
Then, it is seen that the parameter $C$, or $a$ due to
(\ref{a-C}), does appear in (\ref{lw2}), but does not in
(\ref{lc2}). This is a different result obtained from the Lewis
spacetime for the Weyl class taking $b = 0$ and $c = 0$, in which
the spacetime reduces to the static LC spacetime.

    For the van Stockum solution (\ref{vanStockum}), we
find that $V_0, V(r)$ and $G(r)$ are as follows\cite{opher}
\begin{eqnarray}
G(r) &=& 1, \\ V_0 &=& E^2 + 2 \alpha E L - \epsilon,
\\ V (r) &=& \alpha^2 E^2 r^2 + L^2 r^{-2} + K^2
e^{\alpha^2 r^2}.
\end{eqnarray}
In this case, $V(r)$ is non-negative and, in order to have Eq.
(\ref{rdot2}) meaningful for real $r$, we must have $V_0 > 0$,
which is equivalent to
\begin{equation}
E > (\alpha^2 L^2 + \epsilon)^{1/2} -\alpha L.
\end{equation}

    Finally, we note that Eqs. (\ref{soln1_2}), (\ref{soln3_2}),
(\ref{soln4_2}), and (\ref{rdot2}) describe the motion of test
particles in the background of the general cylindrically symmetric
stationary spacetimes. Using these equations, it can easily be
obtained the orbits for the particles, and the effective radial
potential.

\section{Conclusion}

In this paper, we have presented the Dirac analysis of geodesic
equations of the general cylindrically symmetric stationary
spacetimes in explicit form. Using Dirac's theory of constraints
and the covariant Witten-Zuckerman approach we have obtained the
Hamiltonian operators. The results for the symplectic $2$-form
coincide in both of these theories. We note that the original
Lagrangian (\ref{lag1l}) which is second-order is non-degenerate,
and gives second-order geodesic equations which are given in
\ref{apa}. However, the first-order Lagrangian (\ref{lag2}) is
degenerate and produces first-order geodesic equations. Therefore,
if we consider this first-order Lagrangian given by (\ref{lag2}),
then we can easily find the solution of the obtained first-order
geodesic equations of motion for the considered spacetimes. In the
previous section, Sec. \ref{int}, using this degenerate Lagrangian
approach, we have integrated the geodesic equations of motion for
the general cylindrically symmetric stationary spacetimes, and
found some solutions for Lewis, LC, and Van Stockum spacetimes.
Also, for the radial speed of the test particle, we have found a
generic expression given in (\ref{rdot_son}) which is depend on
the solution of Eq. (\ref{eqM}). In some spacial cases, we have
solved the Eq. (\ref{eqM}) and found some solutions for the radial
speed of the test particle.

\section*{Acknowledgements}

This work was done in the {\it Geometry and Integrability}
Research Semester at Feza G\"ursey Institute. I wish to thank
Prof. Dr. Y. Nutku for helpful discussions and I would like to
express profound gratitude to the Feza G\"{u}rsey Institute for
the hospitality.

\appendix

\section{Second-Order Geodesic Equations of Motion}
\label{apa}

For the spacetime (\ref{met}), it follows from the Lagrangian
(\ref{lag1l}) that the geodesic equations of motion (\ref{geoeq})
are
\begin{eqnarray}
& & D\ddot{t} + \frac{\ell f'+k k'}{D}\dot{t}\dot{r} +
\frac{k\ell' - \ell k'}{D} \dot{r} \dot{\phi} = 0,
\label{scndord1} \\& & 2\ddot{r} + e^{-\mu}(f' \dot{t}^2 -
2k'\dot{t}\dot{\phi} -l'\dot{\phi}^2) = 0, \label{scndord2}\\& &
\ddot{z} + \mu' \dot{r} \dot{z} = 0, \quad D \ddot{\phi} + \frac{f
k'-k f'}{D} \dot{f}\dot{r} + \frac{f\ell' + k k'}{D} \dot{r}
\dot{\phi} = 0. \label{scndord3}
\end{eqnarray}

\end{document}